\documentclass[conference]{IEEEtran}
\usepackage{booktabs}
\usepackage{wrapfig}
\usepackage{epsfig}
\usepackage{amssymb}
\usepackage{amsmath, amsfonts}
\usepackage{latexsym, amssymb}
\usepackage{setspace}
\usepackage{longtable}
\usepackage{graphicx}
\usepackage{grffile}
\usepackage{url}
\usepackage{caption}
\usepackage{subfigure}
\usepackage{algorithm}
\usepackage{algorithmic}
\usepackage{multirow}
\usepackage{supertabular}
\usepackage{listings}
\usepackage{color}

\usepackage{courier}

\usepackage{float}
\newenvironment{query}{\begin{tabbing}\makebox[0.0in]{}\=\+\kill}{\end{tabbing}}
\newcommand{\al}[1]{\'#1\\}

\newcommand{\tabincell}[2]{\begin{tabular}{@{}#1@{}}#2\end{tabular}}

\begin{document}
\title{Hardware/Software Co-monitoring}
\author{\IEEEauthorblockN{Li Lei\IEEEauthorrefmark{1}, Kai Cong\IEEEauthorrefmark{2}, Zhenkun Yang\IEEEauthorrefmark{1}, Bo Chen\IEEEauthorrefmark{2}, and Fei Xie\IEEEauthorrefmark{2}}
\IEEEauthorblockA{\IEEEauthorrefmark{1} Intel Labs, Hillsboro, OR 97124}

\IEEEauthorblockA{\IEEEauthorrefmark{2} Portland State University,
Portland, OR 97207}}

% make the title area
\maketitle
%\input{abstract.tex}

%\setcounter{footnote}{0}

%-------------------------------------------------------------------------

%
%%-------------------------------------------------------------------------
\begin{abstract}

Hardware/Software (HW/SW) interfaces, mostly implemented as devices and device drivers, are pervasive in various computer systems. Nowadays HW/SW interfaces typically undergo intensive testing and validation before release, but they are still unreliable and insecure when deployed together with computer systems to end users. Escaped logic bugs, hardware transient failures, and malicious exploits are prevalent in HW/SW interactions, making the entire system vulnerable and unstable.

We present HW/SW co-monitoring, a runtime co-verification approach to detecting failures and malicious exploits in device/driver interactions. Our approach utilizes a formal device model (FDM), a transaction-level model derived from the device specification, to shadow the real device execution. Based on the co-execution of the device and FDM, HW/SW co-monitoring carries out two-tier runtime checking: (1) device checking checks if the device behaviors conform to the FDM behaviors; (2) property checking detects invalid driver commands issued to the device by verifying system properties against driver/device interactions. We have applied HW/SW co-monitoring to five widely-used devices and their Linux drivers, discovering 9 real bugs and vulnerabilities while introducing modest runtime overhead. The results demonstrate the major potential of HW/SW co-monitoring in improving system reliability and security.

\end{abstract}

\section{Introduction} \label{sec: intro}
%\subsection{Motivation}
Computer systems, such as desktops/servers, smart phones, and IoT devices, are pervasive and ever growing. Hardware/Software (HW/SW) interfaces, mostly implemented as devices and device drivers, are central and critical to these systems. For example, about 70\% Linux kernel source code implements device drivers and in the Windows kernel, there are even more drivers as it supports more devices~\cite{Ryzhyk2009, Kadav2012}. Unfortunately HW/SW interfaces are neither reliable nor secure. A majority of system failures are caused by errors in HW/SW interfaces: 85\% of Android kernel (a close fork of Linux kernel) bugs/vulnerabilities are from drivers~\cite{Stoep2016}; studies on Windows support a similar conclusion~\cite{Swift05}. Nowadays assuring the dependability of devices/drivers mainly relies on intensive testing and validation before they are released with computer systems.

Although effective validation can largely improve the quality of devices and drivers, HW/SW interfaces are still facing serious threats after they are deployed to end users.  Generally, there are three major categories of threats at the deployment stage. First, validation cannot discover all the bugs in devices and drivers due to limited amount of time and depth of testing. The uncovered bugs can cause failures over HW/SW interactions at runtime. Second, hardware transient failures are common, particularly in IoT devices and when they are operating under extreme circumstances that were not simulated at the validation stage. Third, recently software malware as well as hardware Trojans have seen significant growth. HW/SW interfaces are vulnerable to the malicious exploits from both hardware and software.

Given the ubiquity and seriousness of such problems, it is critical to develop systematic methods to validate HW/SW interfaces at runtime even after they are released with computer systems. However, the state-of-the-art validation techniques cannot be directly applied at the deployment stage. There are three major challenges. First, testing devices and drivers separately is not sufficient as device/driver interactions are often missed. Second, existing co-verification approaches~\cite{fmsd02} operate against hardware models rather than the real hardware devices which runtime validation needs to deal with. There is a non-trivial gap between hardware models and real devices. Third, hardware devices have limited observability, causing difficulties for runtime co-verification. The techniques used at the validation stage, such as  on-chip monitor~\cite{Abramovici06, Park08}, instrumentation~\cite{Boule06,Hu03,Jos03}, are impractical at the deployment stage due to their heavy weight and intrusive nature. %{ XieYS07, Li2011}

\begin{figure}[h]
\centering
\includegraphics[width=\linewidth]{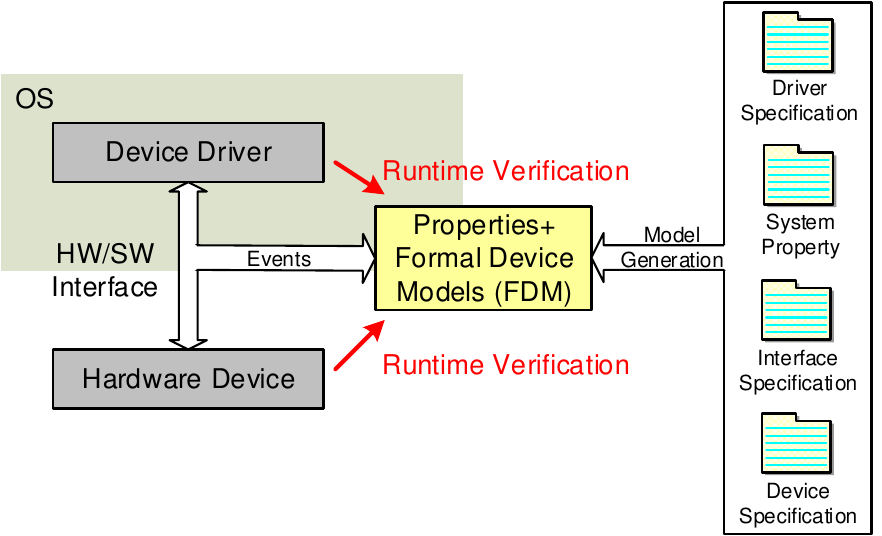}
\caption{HW/SW co-monitoring of device and driver}\label{Fig: comon_1}
\end{figure}

We present {\bf HW/SW co-monitoring}, a runtime co-verification approach to detecting bugs, failures, and malicious exploits over HW/SW interactions. The foundation of our approach is a Formal Device Model (FDM) which is a transaction-level, executable model derived from the device and driver specifications. HW/SW co-monitoring is based on the co-execution of the FDM and device where the FDM shadows the device execution, as shown in Figure~\ref{Fig: comon_1}. With the co-execution, our approach entails three major techniques to realize runtime HW/SW co-verification:
\begin{itemize}
\item Concolic execution of the FDM and driver where the driver is running concretely while the FDM is executed symbolically;
\item Runtime detection of divergence between the device and the FDM, namely device checking;
\item Runtime verification of system properties against the device/driver interaction indirectly over the FDM/driver concolic execution trace.
\end{itemize}

This approach addresses the three challenges mentioned above. First, the device and the driver are verified together through the driver/FDM composition with concolic execution. Second, the FDM shadows the device execution with device checking such that the FDM execution reflects the real device behaviors, bridging the gap between the verification model and the real device. Third, the FDM is executed symbolically while the device internal states are modeled as symbolic values in the FDM. As a result, the FDM execution explores all the possible device internal behaviors, overcoming the limited observability of the device. HW/SW co-monitoring provides a unified solution for detecting and analyzing HW/SW interface defects. By efficiently monitoring the device, the driver, and their interface, it can discover defects across HW/SW interfaces, ranging from hardware transient errors to driver bugs to malicious exploits.

We evaluated our approach on five hardware devices and their Linux drivers. It detected 9 real bugs in both devices and drivers. The overhead is modest: CPU and memory usage increase to 1.2x - 1.6x and 1.1x - 1.4x respecively.  The results demonstrate that hardware/software co-monitoring has major potential in improving system reliability and security.

\medskip
In summary, this paper makes three key contributions:
\medskip
\begin{itemize}
\item We propose HW/SW co-monitoring, the first approach to runtime HW/SW co-verification for detecting failures and malicious exploits over HW/SW interfaces
\item HW/SW co-monitoring integrates several techniques, device checking, concolic execution etc., to resolve the key difficulties of applying HW/SW co-verification techniques at runtime
\item Our approach is able to detect real bugs in commercial device and driver products at the deployment stage
\end{itemize}
\medskip
\noindent {\bf Outline.} The rest of the paper is organized as follows. Section~\ref{Sec: back} presents background concepts. Section~\ref{Sec: overview} overviews the design and implementation of HW/SW co-monitoring. Section~\ref{Sec: secure} discuss the threat model and how our approach can be applied to detect malicious exploits. Section~\ref{Sec: Eva} and Section~\ref{Sec: related} show the evaluation results and related work. Section~\ref{Sec: con} concludes and discusses future work.

\section{Background} \label{Sec: back}
This section presents three concepts that are foundational to our approach: (1) Formal Device Model (FDM), which we use as reference models to monitor hardware devices; (2) symbolic execution, which we use to explore possible FDM behaviors at runtime.

\subsection{Formal Device Model (FDM)} \label{subSec: FDM}

A FDM, proposed by previous work (reference removed for double blind review), is a transaction-level model, specifying HW/SW interfaces and hardware functionalities. It is derived from the device specification and written in a restricted subset of the C language with two key extensions: transaction and non-determinism. %\cite{Li2011}
\begin{itemize}
\item {\em  Transaction.} A FDM focuses on the design logic rather than the implementation details of HW/SW interfaces and hardware functionalities. For example, a data-transfer command is usually processed in multiple clock cycles; however, from the perspective of the software, it is only necessary to describe this command as one hardware state transition. We define a hardware transaction as a hardware state transition in an arbitrarily long but finite sequence of clock cycles. Hardware transactions are atomic from the viewpoint of software.
\medskip
\item {\em Non-determinism.} Hardware devices are concurrent in nature. For example, a network adapter processes driver requests and receives data concurrently. A FDM models this concurrency by using a technique, namely non-deterministic interleaving. Non-deterministic interleaving schedules the concurrent modules (e.g., processing driver requests, receiving data, etc.) with a non-deterministic choice in a loop. While the loop is executed multiple times, the modules are executed in a non-deterministic sequence. The possible effects of hardware concurrency can be captured by the set of hardware states after non-deterministic many executions of the loop. For example, a network adapter concurrently processing a driver request and receiving a packet can be captured by either processing a driver request followed by receiving a packet, or invoking two modules in the reverse order.
\end{itemize}
\begin{figure}[htb]
\begin{center}
\includegraphics[width=\linewidth]{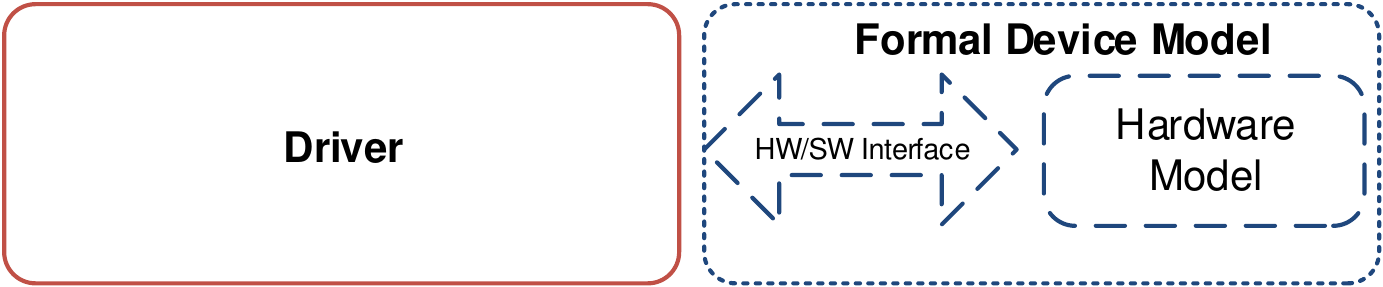}\\
\caption{Structure of FDM with its driver} \label{Fig: FM_COM}
\end{center}
\end{figure}
As Figure~\ref{Fig: FM_COM} shows, a FDM has two components: HW/SW interface modeling the device/driver interactions and hardware model specifying the device functionalities. We illustrate the FDM by using Intel PRO/100 Ethernet adapter (EEPRO100) as an example, which we will use as a running example throughout the paper. As Figure \ref{Fig:FDM-E100} shows, the function {\tt eepro100\_reg\_writel} is a device interface transaction function,, which simulates that the driver writes data to a register of the device. Its parameters {\tt offset} and {\tt value} specify the address of the register and the value written to the register. The function {\tt run\_device} is an internal transaction function with a non-deterministic choice {\tt choice\_device} to schedule the concurrent modules. By invoking {\tt run\_device} non-deterministic many times, the device concurrency is captured by the FDM.
\definecolor{mygreen}{rgb}{0,0.6,0}
\definecolor{mygray}{rgb}{0.5,0.5,0.5}
\definecolor{mymauve}{rgb}{0.58,0,0.82}
\lstset{ %
  backgroundcolor=\color{white},   % choose the background color; you must add \usepackage{color} or \usepackage{xcolor}
  basicstyle=\ttfamily\footnotesize,        % the size of the fonts that are used for the code
 belowskip = 0.02pt,
  aboveskip = 0.02pt,
 abovecaptionskip=0.02pt,
 belowcaptionskip=0.02pt,
  breakatwhitespace=false,         % sets if automatic breaks should only happen at whitespace
  breaklines=true,                 % sets automatic line breaking
  captionpos=b,                    % sets the caption-position to bottom
  commentstyle=\color{mygreen},    % comment style
  deletekeywords={...},            % if you want to delete keywords from the given language
  escapeinside={\%*}{*)},          % if you want to add LaTeX within your code
  extendedchars=true,              % lets you use non-ASCII characters; for 8-bits encodings only, does not work with UTF-8
  frame=single,                    % adds a frame around the code
  keepspaces=true,                 % keeps spaces in text, useful for keeping indentation of code (possibly needs columns=flexible)
  keywordstyle=\color{blue},       % keyword style
  language=C,                 % the language of the code
  morekeywords={dcc_assert, uint32_t, uint8_t},            % if you want to add more keywords to the set
  %numbers=left,                    % where to put the line-numbers; possible values are (none, left, right)
  %numbersep=5pt,                   % how far the line-numbers are from the code
  %numberstyle=\tiny\color{mymauve}, % the style that is used for the line-numbers
  rulecolor=\color{black},         % if not set, the frame-color may be changed on line-breaks within not-black text (e.g. comments (green here))
  showspaces=false,                % show spaces everywhere adding particular underscores; it overrides 'showstringspaces'
  showstringspaces=false,          % underline spaces within strings only
  showtabs=false,                  % show tabs within strings adding particular underscores
%  stepnumber=1,                    % the step between two line-numbers. If it's 1, each line will be numbered
  stringstyle=\color{mymauve},     % string literal style
  tabsize=2,                       % sets default tabsize to 2 spaces
%  title=\lstname                   % show the filename of files included with \lstinputlisting; also try caption instead of title
}
\begin{figure}[htb]
   %   \begin{lstlisting}[language=C]
  % \footnotesize
%\scriptsize
         \begin{lstlisting} [linewidth=\linewidth]
void eepro100_reg_writel(uint32_t offset, uint32_t value)
{
     switch (offset){
	      case EEPRO100_CTRL:
             eepro100_write_ctrl(value);
             break;
	      case EEPRO100_MDIC:
             eepro100_write_mdic (value);
             break;
	         ... ...
	      default:
             break;
     }
}

void run_device()
{
    ......
    // Non-deterministic choice
    switch(choice_device()){
     case 1:
        eepro100_run_command(s); break;
     case 2:
        eepro100_receive(nc, buf, size);
        break;
        ......
    }
}
          \end{lstlisting}
\caption{Excerpts from EEPRO100 FDM.} \label{Fig:FDM-E100}
\end{figure}

\subsection{Symbolic Execution}
Symbolic execution \cite{King76} executes a program with symbolic values as inputs instead of concrete ones and represents the values of program variables as symbolic expressions. Consequently, the outputs computed by the program are expressed as functions of input symbolic values. The symbolic state of a program includes the symbolic values of program variables, a path condition, and a program counter. The path condition is a Boolean expression over the symbolic inputs; it accumulates constraints which the inputs must satisfy for the symbolic execution to follow the particular associated path. The program counter points to the next statement to be executed. A symbolic execution tree captures the paths explored by the symbolic execution of a program: the nodes represent the symbolic program states and the arcs represent the state transitions.

\begin{figure}[htb]
\begin{center}
    \begin{minipage}[t]{2.3in}

      \scriptsize
      \begin{query}
\al{}
\al{}
\al{}
	\al{~~~~~~~~~~~~~~~~void foo(int a,  int b) \{}
	\al{1:~~~~~~~~~~~~~~~~~~~if ( a $>$ 5)}
	\al{2:~~~~~~~~~~~~~~~~~~~~~~b++;}
	\al{3:~~~~~~~~~~~~~~~~~~~else}
	\al{4:~~~~~~~~~~~~~~~~~~~~~~b=b-1;}
	\al{~~~~~~~~~~~~~~~~\}}
	\end{query}
	\end{minipage}
	\quad \quad
\begin{minipage}[t]{1.3in}
      \begin{query}
	\al{}
	\includegraphics[height=3cm,width=4cm]{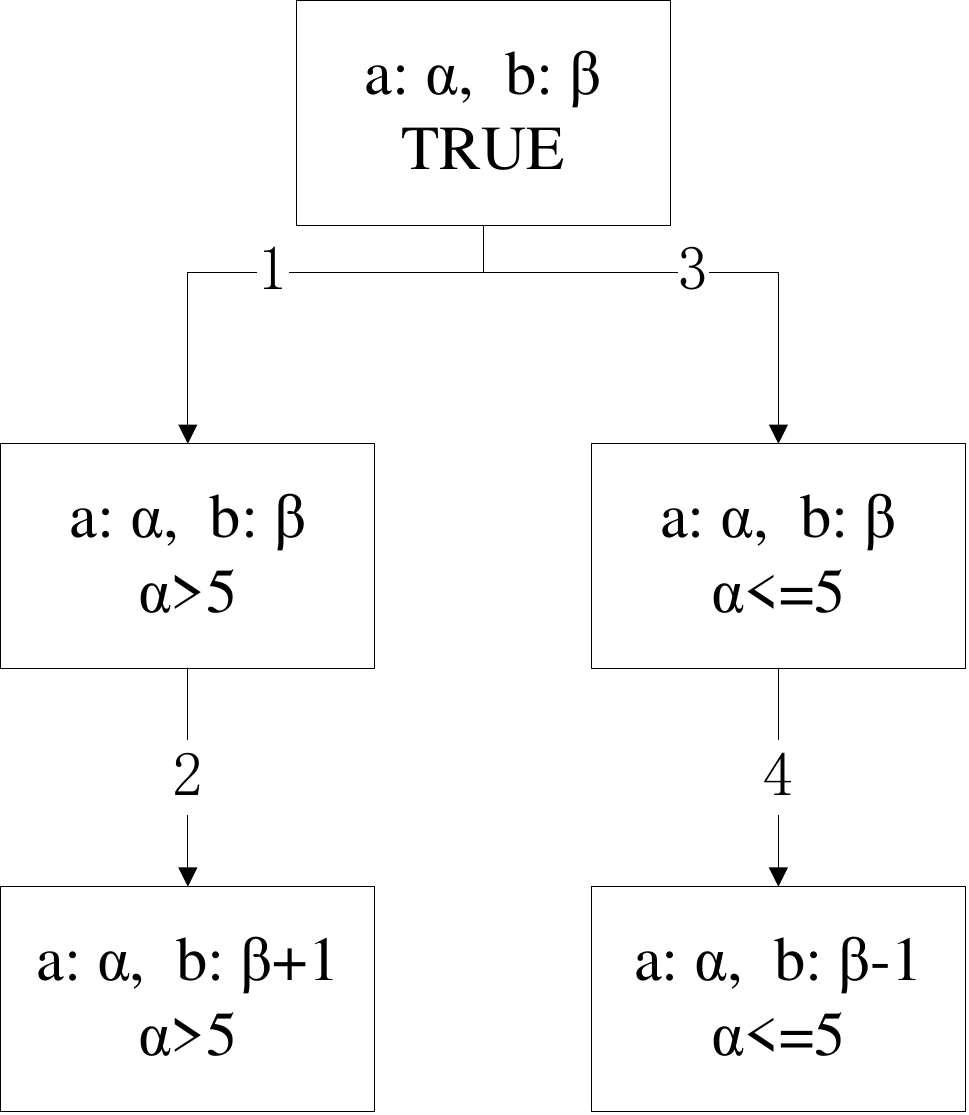}
	\end{query}
	\end{minipage}
 	\end{center}
\caption{Example of symbolic execution.}\label{Fig: Ex-Sym}
\end{figure}

We use the program in Figure \ref {Fig: Ex-Sym} to illustrate how symbolic execution is conducted. At the entry, $a$ and $b$ have symbolic values $\alpha$ and $\beta$, respectively, the path condition is {\tt TRUE}, and the program counter is 1. At the branching point, the path condition is updated with conditions on the inputs to select between the two alternative paths. At an assignment statement, the symbolic value of the relevant variable is updated.

\section{Design and Implementations} \label{Sec: overview}
\subsection{Overview}
\begin{figure}[htb]
\begin{center}
\includegraphics[width=\linewidth]{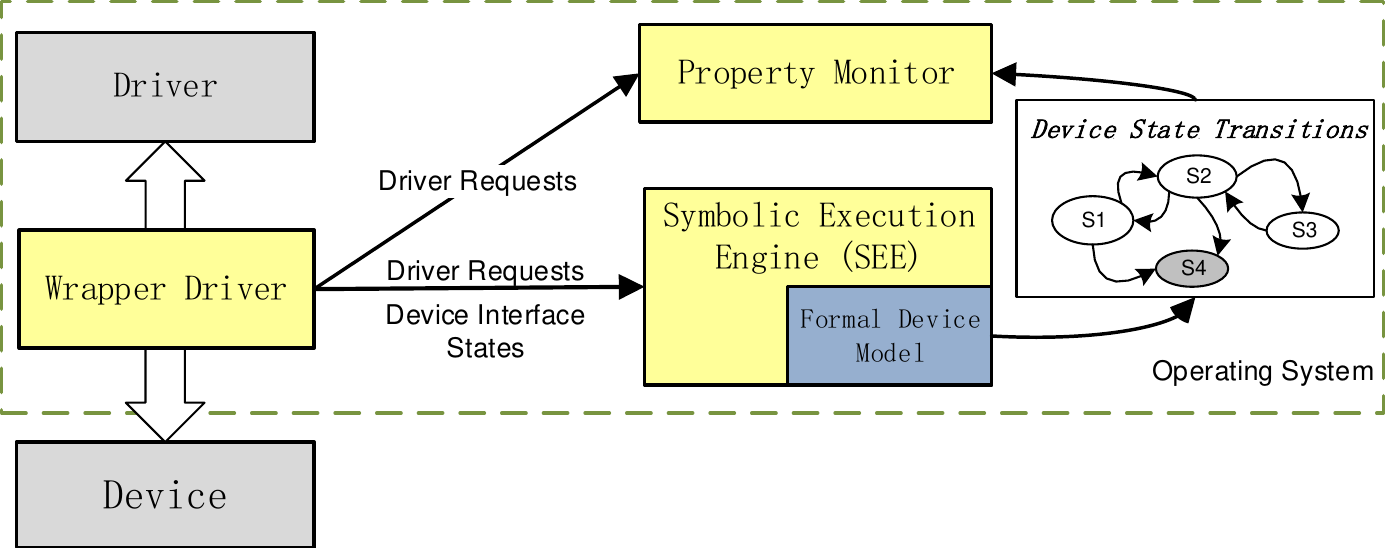}
\end{center}
\caption{The architecture of HW/SW co-monitoring}\label{Fig:FM_3}
\end{figure}

The HW/SW co-monitoring framework includes three components: a wrapper driver, a symbolic execution environment (SEE), and a property monitor (PM), shown in Figure~\ref{Fig:FM_3}. The wrapper driver captures the device/driver interactions, including the driver requests issued to the device and the device interface states exposed to the driver. The SEE symbolically executes the FDM by taking the driver request sequence as inputs, realizing the co-execution with the device.  With the co-execution, the SEE conducts device checking to ensure that the FDM shadows the device execution. The PM monitors the device/driver interactions by enforcing system properties which the device/driver interactions should follow.

\noindent{\bf Device Checking.} Device checking detects the divergence between the FDM and the device at runtime. SEE employs a technique, conformance checking, which is originally applied to check the conformance between a device and its virtual prototype in previous work. We extend the conformance checking to check the conformance between a device and its FDM. Non-conformance between the FDM and device indicates the logic errors or exploits occurring in the device interface. With the conformance between the FDM and device, the FDM shadows the device execution, exposing the device state transition to the PM for property checking.
\medskip

\noindent{\bf Property Checking.} Property checking verifies if the device/driver interactions follow certain system properties. PM takes the driver requests and the device state transitions exposed through device checking as its inputs, checks if the enforced properties are violated in the device state transitions under the driver requests. A property violation typically indicates the error or malicious exploits from the driver.

\subsection{Preliminaries} \label{subSec: Prelimininary}
Based on the notion of conformance checking over devices and virtual prototypes proposed in previous work (reference removed for double blind review), we define conformance between the FDM and the device. A hardware device state is composited by a set of interface state variables (e.g., registers) denoted as $R_I$ and a set of internal state variables denoted as $R_N$. As a result, we define a device state as Definition~\ref{Def: silicon state}. Since at runtime the device internal state cannot be observed, we assign symbolic values to the device internal state variables. \medskip
\newtheorem{definition}{Definition}
\begin{definition}\label{Def: silicon state}
A {\bf device state} is denoted as $S=\langle S_{I}, S_{N} \rangle$ where $S_{I}$ is the assignments to variables in $R_I$ and $S_{N}$ is the symbolic assignments to variables in $R_N$.
\end{definition}\medskip

Correspondingly, we define a FDM state following the same structure of the device state as follows. \medskip

\begin{definition}\label{Def: virtual state}
A {\bf FDM state} is denoted as $V= \langle V_{I}, V_{N} \rangle$ where $V_{I}$ is the device interface state, i.e., the assignments to variables in $R_I$ and $V_{N}$ is the device internal state, i.e., the assignments to variables in $R_N$.
\end{definition}\medskip

In device checking, the FDM always starts execution from a same state as the device, then co-execution with the device by taking the same sequence of driver requests. As a result, in a FDM state, some internal variables in $R_N$ could have symbolic values as well.

A {\bf concrete state} is a state whose state variable values are all concrete. A {\bf symbolic state} is a state some of whose state variable values are symbolic and there can be constraints on these symbolic values. A symbolic state can be viewed as a set of concrete states.

Both device and FDM states are symbolic values and they can be viewed as a set of concrete states. We define the conformance between a FDM state and the device state as follows. \medskip

\begin{definition}[State Conformance]\label{Def: stateconform}
A device state $S$ conforms to a FDM state $V$ if $S \cap V \neq \emptyset$.
\end{definition} \medskip

$V$ and $S$ represent all the possible concrete states that the FDM and the device may have respectively. The condition $S \cap V = \emptyset$ indicates that the FDM state and the device state cannot be the same, i.e., the device and FDM executions diverge at runtime.

\subsection{Wrapper Driver} \label{subsec: wrapper}
The wrapper driver captures: (1) the driver request issued to the device; (2) the device interface state before the driver request issued. When the driver issues a new request, the wrapper driver captures and sends such a state-request pair to the SEE. We denote such a state-request pair as $\langle S_I, \alpha \rangle$ where $S_I$ is the device interface state and $\alpha$ is the current driver request.

In practice, capturing device interface states has a challenge: a peripheral device often has a large range of registers. Capturing all the registers heavily degrades the system performance, which is infeasible. To overcome this difficulty, we employ a technique, register sampling, to selectively capture the hardware interface registers.

\medskip
\noindent {\bf Register Sampling.} The interface state of the device can be divided into two types of registers: (1) the registers engaged with device normal functionalities; (2) reserved registers. The size of the first class of registers is typically small while the second is large. In HW/SW co-monitoring, the wrapper driver need to capture both of them. The functional registers are essential to simulate the device behaviors on the FDM while the reserved registers are also interesting to us as they can be used for defining abnormal behaviors and malicious attacks. For example, the reserved registers can be used to place the executable code for code injection attacks (see details in Section~\ref{subsec: HA}).

Our capturing method captures all the registers which are essential to the device functionalities. To capture reserved registers, we design a sampling method. It samples the reserved registers rather than capturing all of them: from the beginning to the end of the register range, it captures one reserved register every few registers. This sampling method is effective in detecting malicious exploits due to two reasons: (1) for the attacks using a large range of reserved registers, sampling can easily hit part of it; (2) for the attacks using a smaller range, we increase the chance of detection by sliding the sample windows every time we capture. The evaluation shows that this method discovered several incorrect values of reserved registers (see details in Section~\ref{subSec: bugs}).

%\subsubsection{Sampling Reserved Registers}
%Selective capturing helps capture the registers essential to the device functionalities. However, directly adopting this method does not fully meet our requirements for monitoring the device. The goal of HW/SW co-monitoring is to detect not only logic bugs, but also the undefined or abnormal, even malicious behaviors over device/driver interfaces. Therefore, we also need to capture the registers which are not essential to the functionality such as reserved registers as these registers can be used for malicious attacks.
%
%Nevertheless the size of reserved registers is often too large to capture. We develop a method which samples the reserved registers rather than capturing all of them:

\subsection{Device Checking} \label{subsec: rdc}
In device checking, symbolic execution is used to enable the co-execution of the FDM/device since the FDM has non-deterministic choices and its internal variables assigned with symbolic values. SEE symbolically executes the FDM while continuously taking the state-request pairs from the wrapper driver.  Algorithm \ref{Alg: tReplay} presents the work flow of device checking. It takes a FDM $\mathcal{M}$ as inputs and the functions in Algorithm~\ref{Alg: tReplay} are described as follows.
%\begin{enumerate}
%\item SEE gets the first request-state pair and intiaize FDM state $V$ with the device interface state $S_I$;
%\item SEE symbolically executes the FDM with the driver request $\alpha$ and the FDM state $V$;
%\item SEE checks the conformance between the set of possible FDM next states $G$ and the device next state $S^\prime$;
%\item If the device conforms to the FDM, we construct the next FDM state and assign it to $V$, go to step 2. Otherwise, we report a device error.
%\end{enumerate}
\renewcommand{\algorithmicforall}{\textbf{for each}}
\begin{algorithm} [htb]
\begin{algorithmic}[1]
\caption{Device\_Checking($\mathcal{M}$)} \label{Alg: tReplay}
\STATE{$ \langle S_I, \alpha \rangle \leftarrow receive\_state\_request()$}
\STATE{$S \leftarrow construct\_device\_state(S_I)$}
\STATE{/*Initialize FDM state $V$ to be device state $S$*/}
\STATE{$V \leftarrow S$}
%\DO{$D_k$ of $T^\prime$}
\WHILE{$\alpha \neq $ {\tt NULL}}
\STATE{/*Symbolically execute FDM taking $\alpha$ at $V$ state.*/}
\STATE{$G \leftarrow sym\_exec(\mathcal{M}, V, \alpha)$}
\STATE{$ \langle S_I^\prime, \alpha^\prime \rangle \leftarrow receive\_state\_request()$}
\STATE{$S ^\prime \leftarrow construct\_device\_state(S_I^\prime)$}
\STATE{$H \leftarrow conformance\_check(G, S^\prime)$}
\IF{$(H == \emptyset)$}
\STATE{$report\_device\_error()$}
\STATE{$abort()$}
\ENDIF
\STATE{$V^\prime \leftarrow construct\_next\_State(H)$}
\STATE{$V \leftarrow V^\prime$}
\STATE{$ \alpha \leftarrow \alpha^\prime$}
\ENDWHILE
\end{algorithmic}
\end{algorithm}
\begin{enumerate}
\item{\em Receiving Requests.} Function $receive\_state\_request()$ is invoked to wait for and receive a state-request pair $\langle S_I, \alpha \rangle$ from the wrapper driver.
\item{\em Device State Construction.} Given a device interface state $S_I$, based on Definition~\ref{Def: silicon state}, $construct\_device\_state$ constructs a device state $S = \langle S_I, S_N \rangle$ where state variables in $S_N$ are assigned symbolic values.

\item{\em Symbolic Execution.} Function $sym\_exec$ symbolically executes $\mathcal{M}$ and generates a set of FDM states, denoted as $G$.

\item{\em Conformance Checking.} Symbolic execution of $\mathcal{M}$ may produce a set of FDM states $G=\{g_i\mid0 \leq i \leq n\}$. The next device state received from the wrapper driver is denoted as $S^\prime$. Function $conformance\_check$ checks the conformance between the device and FDM. Definition~\ref{Define: Conformance} defines their conformance based on $G$ and $S^\prime$. We explain the definition under Definition~\ref{Define: Conformance}. Function $conformance\_check$ returns a subset of $G$, each FDM state of which conforms to the device state, denoted as $H =\{h_i \mid 0 \leq i \leq m\}$.

\item{\em Device Error Report.} If $H$ is empty, no conforming FDM state is produced, i.e., the FDM does not conform to the device at driver request $\alpha$, function $report\_device\_error$ is invoked to record the device error, including the FDM execution trace, the driver request, and the state variables of the device, which are not equal to the state variables of the FDM.

\item{\em Next State Construction.} If $H$ is not empty, the FDM and the device conform to each other at $\alpha$. Based on $H =\{h_i \mid 0 \leq i \leq m\}$, function $ construct\_next\_State$ constructs the next FDM state $V^\prime$ as follows: $V^\prime = \bigcup\limits_{i=1}^m (h_i \cap S^\prime)$.
\end{enumerate}

\medskip
\begin{definition}[Device Conformance]\label {Define: Conformance}
Given the set of FDM states $G$=$\{g_i \mid 0 \leq i \leq n\}$ and the device state $S^\prime$, the device conforms to the FDM at $\alpha$ if $\exists g_i \in G$ where $0 \leq i \leq n$, $S^\prime \cap g_i \neq \emptyset$.
\end{definition}

\medskip
The set $G$ and the device state $S^\prime$ represent all the possible concrete states the FDM and the device can have after processing $\alpha$ respectively. Thereby the empty intersection of $G$ and $S^\prime$ indicates there is no possibility such that the device and FDM have the same state. With the empty intersection, device checking can ensure that the device and the FDM are divergent at runtime. We essentially adopt a conservative strategy for detecting device bugs, i.e., we pursue soundness rather than completeness in our approach, as our monitoring framework is deployed to the end users, which typically desires minimum false positives.

\subsection{Property Checking} \label{SubSec: PM}
Through device checking, the FDM shadows the device execution and exposes the device state transitions to PM. PM verifies if the enforced system properties hold over the device state transitions triggered by the driver requests. A property violation indicates the invalid or malformed driver inputs to the device interface registers.
\medskip
\subsubsection{Co-verification with Properties}
According to Section~\ref{subsec: rdc}, under a single driver request, the FDM is symbolically executed, exploring a set of FDM paths, each of which can be represented by a sequence of device state transitions. We denote the set of explored paths as $P$ and the driver request as $\alpha$, Definition~\ref{Define: pm} defines when a property is violated during device/driver interactions.
\medskip
\begin{definition}[Property violations]\label {Define: pm}
{ Given a property $\psi$, a set of paths $P = \{p_i \mid 0 \leq i \leq n\}$ explored under a driver request $\alpha$, $\psi$ is violated under $\alpha$ if $\forall p_i \in P$, $\neg \psi$ is reachable on $p_i$.}
\end{definition}
\medskip
Similar as device checking, $P$ represents all possible device behaviors under the current driver request $\alpha$. Only if all of these possible behaviors lead to the violation of the property $\psi$, PM can ensure there is an invalid driver request triggering the violation.

\subsubsection{Implementation of Property Checking}
\medskip
PM enforces system properties from two major categories: {\em access control property} and {\em protocol property}. An access control property specifies that the driver should follow the access restrictions when accessing the device registers. A protocol property specifies that the device/driver should follow HW/SW interface protocols defined in device specifications. Property 1 is an access control property and Property 2 which is specified in EERPO100 specification is a protocol property. \medskip

\noindent { {\bf Property 1}: The driver should never writes to a read-only register} \medskip

\noindent { {\bf Property 2}: If the device Command Unit (CU) is not in {\tt SUSPENDED} status, the driver cannot send {\tt RESUME} to the device.} \medskip

PM automatically enforces access control properties at runtime. By monitoring the driver requests issued to the device, PM checks if the driver requests violate any access restrictions which are defined in the attributes of the registers of the FDM. The protocol properties are typically related to the state transitions. Therefore, we express them in the term of assertions and instrument assertions into the FDM.  Figure~\ref{Fig: assert} shows how to specify Property 2 with assertions in the FDM by using the provided function $comon\_assert$. According to Definition~\ref{Define: pm}, PM reports a property violation when the assertion fails on all the path explored by symbolic execution of the FDM.

\begin{figure}[htb]
\centering
      \begin{lstlisting}
static void
eepro100_cu_command( EEPRO100State* s, uint8_t val)
{
    ......
    // Assertion to enforce the property
    if (s->cu_state != CU_STATE_SUSPENDED)
   		comon_assert(val != CU_CMD_RESUME);
    ......
}
 \end{lstlisting}
\caption{Assertions instrumented in EEPRO100 FDM} \label{Fig: assert}
\end{figure}

\subsection{Implementation Details}
We implement HW/SW co-monitoring framework on Linux. Nevertheless the similar design can be implemented on Windows or other platforms as well. The wrapper driver is implemented as a loadable kernel module, which intercepts the target driver inputs to the target device and captures device interface states. As a standard Linux driver typically calls Linux kernel functions to manipulate its device. For instance, a driver writes values to device registers by calling Linux kernel functions such as {\tt writel}, {\tt writeb}. The wrapper driver hooks these kernel functions. As a result, when the driver calls the kernel functions to issue requests to the device, the wrapper driver is invoked to record and the driver requests as well as interface registers.

We build SEE on the top of KLEE, a widely-used open-source symbolic execution
frame [4]. KLEE exercises C/C++ based software programs through symbolic execution by taking the LLVM-compiled bitcode of the program as its input. To enable runtime monitoring, the FDM is compiled with LLVM compiler~\cite{Lattner2004} to llvm bit code offline and running online inside modified KLEE. We extend KLEE by integrating our two-tier checking and the module communicating with the wrapper driver running inside the kernel. %We modify KLEE in two aspects. First, we set the loop bounds during symbolic execution. Second, we realize our own module for conformance checking.

\section{Applications in Security} \label{Sec: secure}
Our approach can detect not only logic bugs but also malicious exploits across device/driver interfaces. This section discusses the threat model of our approach by elaborating and classifying the malicious exploits occurring over device/driver interfaces, and shows how HW/SW co-monitoring can catch these exploits at runtime.
\subsection{Threat Model} \label{subsec: threat}
In our threat model, a malicious exploit can be launched either from the device or the driver and attack another side of the device/driver interface. We categorize these exploits into software attacks where malicious software exploits the hardware interface and hardware attacks where malicious hardware exploits the software interface.
\medskip
\subsubsection{Software Attacks}
Peripheral devices expose a wide range interface to the device drivers, i.e., large range of interface registers and DMA memory. Hostile software, e.g., malware or rootkits can manipulate the exposed hardware interfaces in illegal ways, such as writing arbitrary values to the interface registers. The devices are vulnerable to these exploits as device/driver interactions always need to follow certain protocols. The Violations of these protocols can easily drive the device into unknown states, causing denial-of-service (DoS) or even more serious attacks. For example, if the eepro100 driver intensively sends {\tt CU\_CMD\_RESUME} commands to device CU which is not at {\tt CU\_STATE\_SUSPENDED}, violating Property 2 in Section~\ref{SubSec: PM}, the device will be driven to an unresponsive state such that normal functional operations over device/driver are suspended. We demonstrate that we successfully attack the e100 device by violating such a property (See Section~\ref{subsec: att-det} for details).
\medskip
\subsubsection{Hardware Attacks} \label{subsec: HA}
Software interface exposed to hardware is also vulnerable. Generally the device driver takes device inputs from device interface registers and the shared memory in RAM through DMA. A malicious device can always feed invalid even malformed values through those the shared interfaces. By exploiting the driver vulnerabilities with the malformed inputs,  the attacker can articulate the driver to unresponsive states or crashes, or even gain the control of the entire system. We illustrated these attacks in details. \medskip

\begin{figure}[h]
   	\footnotesize
     \begin{lstlisting}

  while (ioread16(ioaddr + Wn7_MasterStatus)
               & 0x8000)
          ;
     ...
          \end{lstlisting}
\caption{A snippet from Linux 3c59x driver} \label{Fig:vuldriver}
\end{figure}

\noindent {\bf Logic Bug Exploitation.} There often exists logic bugs in the driver code which interacts with the device such as improper handling of device inputs. Hostile devices can exploit these bugs with careful-crafted inputs, to hang or crash the driver, leading to denial-of-services attacks. Figure~\ref{Fig:vuldriver} shows a real bug existing in Linux 3c959x driver, which is illustrated in~\cite{Kadav09}. Function ``ioread16'' reads data from the device. If the value read from the device is improper, the driver will loop forever. The malicious attacker inside the device could feed such an improper value to make the driver to loop, causing the device/driver stack even the entire system out of services.

\medskip
\lstset{
  numbers = right,
  numberstyle=\tiny\color{mygray}
 }

\begin{figure}[htb]
\
     \begin{lstlisting}[escapechar=!]
/*
 * Handle raw report as sent by device
 */
static int
picolcd_raw_event(struct hid_device *hdev,
                  struct hid_report *report,
                  u8 *raw_data, int size) {
		...
	if (report->id == REPORT_KEY_STATE) {
		...
	} else if (report->id == REPORT_IR_DATA) {
		...
	} else {
		...
		memcpy(data->pending->raw\_data, raw\_data+1, size-1);
		...
	...
}
		\end{lstlisting}
\caption{A code snippet of Linux picoLCD driver} \label{Fig: pdriver}
\end{figure}

\begin{figure*}[t]
\includegraphics[width=\linewidth]{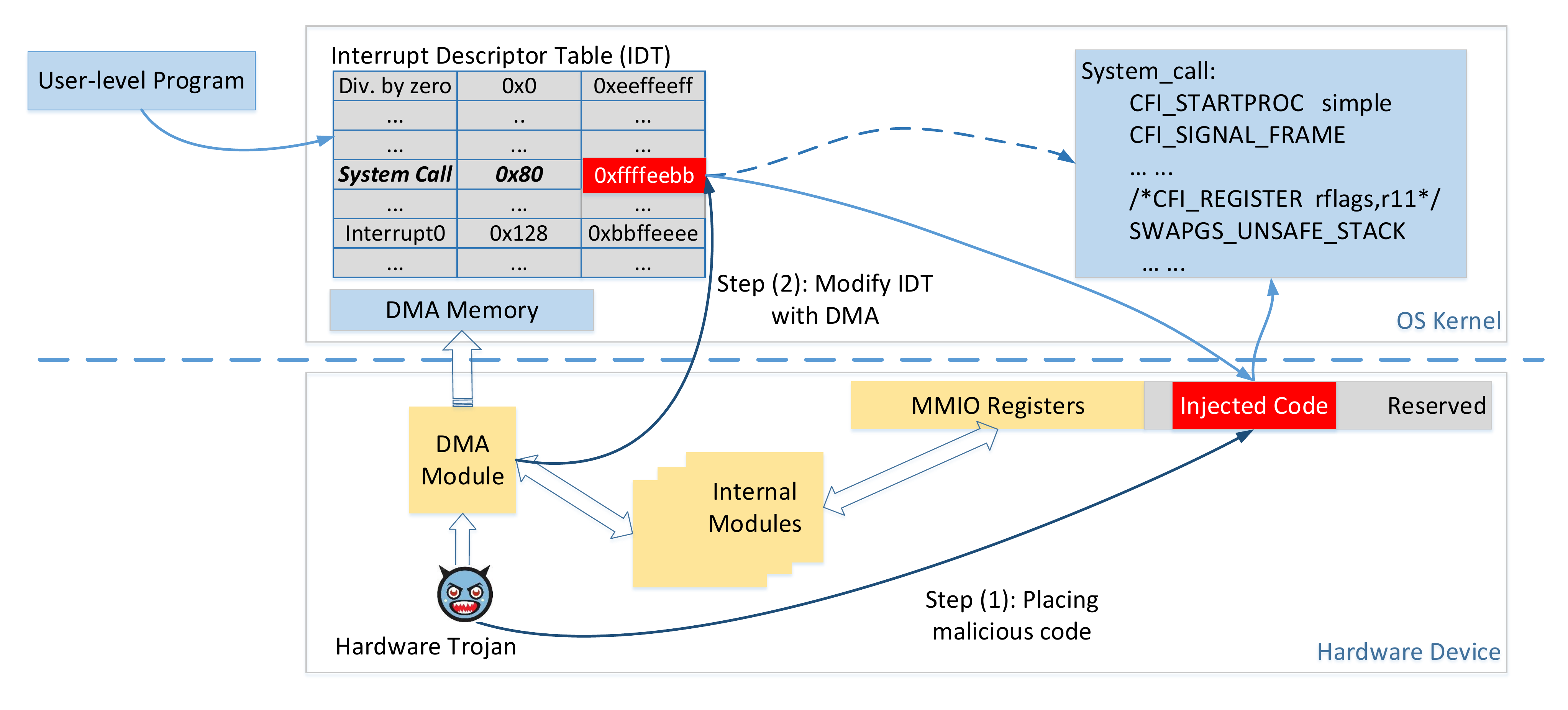}
\caption{A hardware Trojan (HT)-based attack scenario with two steps: (1) HT places injected code in reserved registers; (2) HT hooks system call handlers through DMA attacks  }\label{Fig: Att}
\end{figure*}

\noindent {\bf Software Vulnerability Exploitation.} Security vulnerabilities are common in drivers. Some vulnerabilities can be exploited by device inputs. For instance, a device driver often conducts many memory operations to copy data and commands from the device. Some vulnerable memory operations such as a {\tt memcpy} with buffer overflow can be exploited by malformed device inputs, leading to memory corruption. The attacker can further leverage these corruptions to launch various attacks for denial-of-services,  privileged escalation, and so on. We present a discovered vulnerability in a real driver~\cite{CVE-2014-3186} as an example. Figure~\ref{Fig: pdriver} illustrates a code snippet of the PicoLCD HID device driver. With the code at line 15, the driver copies the raw report data from the device to the driver main memory via the function {\tt memcpy}. The buffer {\tt raw\_data} as well as the copy size {\tt size} are fully controlled by the display device. Since the driver doesn't check the bound of the buffer before copying the data, a malicious device can easily cause buffer overflow and launch attacks thereafter by providing a malicious-crafted raw report data with larger size. There are more reported memory corruption bugs which could be exploited by malformed device inputs with similar methods, causing serious consequences~\cite{CVE-2014-3182, CVE-2014-6410, CVE-2013-2897, CVE-2013-1860}.

\medskip
\noindent {\bf Code Injection Attack.} Modern devices typically have large ranges of interface registers which their device drivers can access. For example, Intel e1000 Ethernet has around 8KB interface registers. The DMA-based shared memory interfaces between devices and drivers are even much larger. Both interface registers and DMA-based memory could be leveraged for code injection attack by malicious devices, i.e., an advisory could place its malicious code into the device interface memory and execute the code by exploiting software vulnerabilities. Several work~\cite{Piegdon2006, Breuk2012} demonstrated that code injection can be combined with a {\em DMA attack}: malicious hardware can modify critical OS kernel structure via illegal DMA accesses and eventually execute the binary code injected into shared interface between the device and driver.

\medskip
\noindent {\bf Example.} We use a DMA-based attack as a concrete example to show how a code injection attack is carried out through updating device interface registers. Figure~\ref{Fig: Att} shows the work flow of this attacking scenario. In this example, a hardware Trojan in an Ethernet adapter takes over the OS by injecting code through DMA attacks. There are two steps:  (1) placing the code to be executed in MMIO reserved registers; (2) hooking system calls by modifying the Interrupt Descriptor Table (IDT) through DMA. As a result, when a system call is executed, the injected code is executed first.  Moreover, injected code is executed in the OS kernel space which has root privilege, causing privilege escalation attacks.

\subsection{Detecting Malicious Attacks}
HW/SW co-monitoring can detect the malicious exploits by effectively monitoring HW/SW interactions. We show how our approach detects each type of attacks listed above. \medskip

\noindent {\bf Detecting Software Attacks.} The property monitor in our framework can help detect software attacks to hardware. In the property monitor, a set of system properties and security policies are enforced. By verifying these properties at runtime, malformed driver commands which violate the properties can be discovered.\medskip

\noindent {\bf Detecting Hardware Attacks.} The hardware attacks listed above generally rely on giving software driver invalid and malformed inputs from hardware devices either through device interface registers, or DMA shared memory. Since the FDM models the normal correct behaviors of the device. Through device checking, any invalid and malformed inputs on device registers and DMA interfaces would lead to interface inconsistencies between the FDM and the device. As a result, by enforcing the device checking, malformed input exploiting driver bugs and vulnerabilities will be   detected at runtime. For code injection attacks, as the executable code is typically injected into hardware interfaces such as interface registers or DMA interfaces, the device checking can detect this abnormal values in the interface by checking the conformance between the FDM and the device. %For the denial-of-service caused by the invalid hardware inputs, i.e., invalid register values, the device checking can catch such incorrect register values by detecting the interface inconsistencies between the device and the FDM.
\medskip

In this section, we elaborate on several potential malicious attacks across HW/SW interfaces. Nevertheless, our approach is not limited in only detecting these attacks. By effectively monitoring HW/SW interfaces, HW/SW co-monitoring is able to discover most of exploits as long as their abnormal behaviors eventually appear on HW/SW interfaces. \medskip

\section{Evaluation} \label{Sec: Eva}
This section evaluates our approach from three aspects. First, we present several real device and driver bugs discovered by HW/SW co-monitoring, demonstrating our approach is effective in catching bugs at runtime. Second, we simulate several types of malicious attacks across HW/SW interfaces and run our co-monitoring framework over that, to demonstrate our approach is effective in catching malicious exploits as well. Third, we evaluate the overhead introduced by our framework.

\subsection{Experiment Setup}
We have performed our experiments on a workstation of Ubuntu Linux OS with 64-bit kernel version 2.6.38. We applied our approach to five Ethernet cards and their FDMs. The FDM and driver size, measured in Lines of Code (LoC), are summarized in Table~\ref{Tab: Dev}.
\begin{table}[htbp]
\centering
 \scriptsize
\caption{Summary of devices for case studies}\label{Tab: Dev}
\begin{tabular}{c|c|c|c} \hline
\tabincell{c}{\bf Devices} & \tabincell{c}{\bf FDM \\ \bf Size (LoC)}& \tabincell{c}{\bf Driver \\ \bf Size (LoC)} & \tabincell{c}{\bf Basic Description } \\\hline
RealTek rtl8139 & 1411 & 7001 & RealTek 10/100M NIC \\\hline
Intel eepro100 & 1032 & 7578 & Intel 10/100M NIC \\\hline
Intel e1000 & 1632 & 15118  & Intel Gigabit NIC\\\hline
Broadcom bcm5751 & 2103 & 12559 & Broadcom Gigabit NIC \\\hline
Intel x520 & 2501 & 21233 & Intel 10 Gigabit NIC \\\hline
\end{tabular}
\end{table}
\subsection{Bug Detection} \label{subSec: bugs}
Our approach detects 9 real bugs in both devices and drivers, shown in Table~\ref{Tab: Bugs}. The driver bugs are related to violation of access control properties and the device bugs are unexpected updates to reserved registers. From the reliability perspective, these bugs might not cause malfunction of the system as they are not part of the normal logic. Nevertheless, from the security perspective, all these violations can be potential malicious exploits, for example, invalid driver inputs, e.g., writing to read-only registers, can be either exploiting hardware interface vulnerabilities or cooperatively working with hardware to launch new attacks such as Stuxnet~\cite{Stuxnet}. The device bugs such as updating reserved registers can be the forms of various attacks as well, for example, code injection attacks might use reserved registers to place the code, presenting the exact same phenomenon as the bugs we discovered. Discovering these non-trivial bugs demonstrates HW/SW co-monitoring is effective for detecting not only logic bugs but also the transient failures as well as the exploits across HW/SW boundaries.

\begin{table}[!h]
\begin{center}
 \scriptsize
\caption{Summary of detected bugs} \label {Tab: Bugs}
\begin{tabular}{c|l|c|c} \hline
 \tabincell{c}{\bf No.}  &\tabincell{c}{\bf Description} & \tabincell{c}{\bf Dev./Drv.} & \tabincell{c}{\bf Num.}\\\hline
 1 & Driver writes to a read-only register. & eepro100 & 1 \\\hline
 2 & Driver updates the reserved register bits & e1000 & 2 \\\hline
3 & Device updates reserved registers & eepro100, e1000, x520 & 6\\\hline
\end{tabular}
\end{center}
\end{table}

\begin{table*}[!t]
\centering
 \scriptsize
\caption{Summary of Software Attack Injection}\label{Tab: Property}
\begin{tabular}{|c|c|c|} \hline
\tabincell{c}{\bf Driver} & \tabincell{c}{\bf Property}& \tabincell{c}{\bf Consequences } \\\hline
rtl8139 & \tabincell{l}{ The driver should not start new transaction when another transaction is in progress} & Device hangs \\\hline
eepro100 & \tabincell{l}{The driver should not issue START while  the device is working already} & \tabincell{c}{Device or even system hangs} \\\hline
e1000 & \tabincell{l}{The driver should not issue command when MDIC is not clear}& Device hangs\\\hline
bcm5751 &\tabincell{l}{The driver should not start new  EEPROM transaction when previous update is not finished}  & Device hangs \\\hline
x520 & \tabincell{l}{The driver should not issue command when MDIC is not clear}& Device hangs\\\hline
\end{tabular}
\end{table*}

\begin{figure*}[!t]
\begin{center}
\includegraphics[height=5cm]{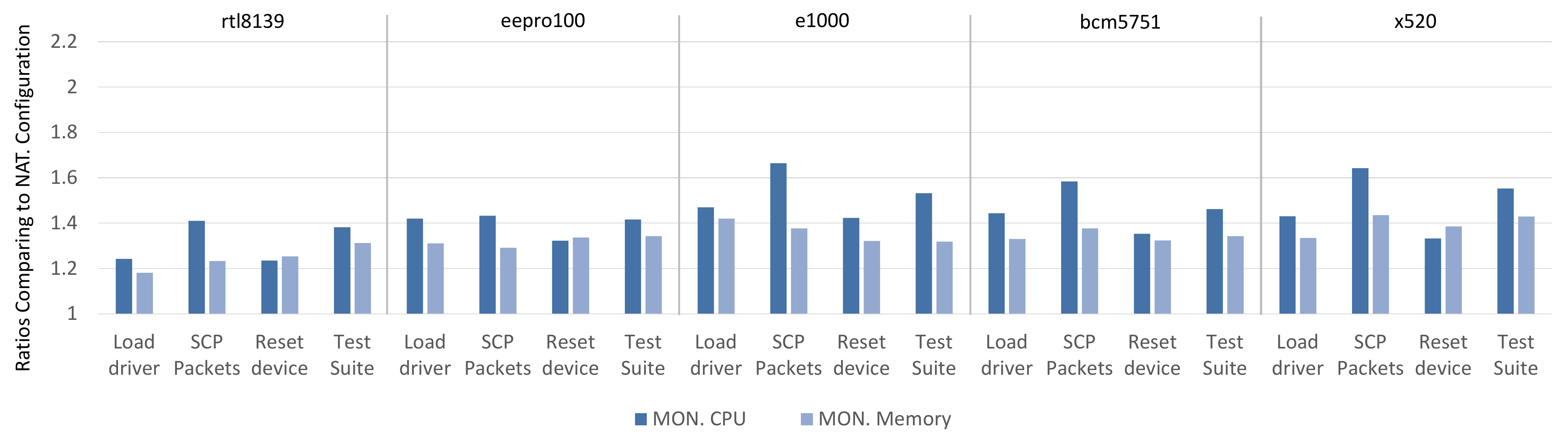}
\caption{CPU and memory usages of test cases under MON configurations. The CPU and memory usages under the NAT configuration are normalized to 1.} \label{Fig:PER}
\end{center}
\end{figure*}

\subsection{Malicious Exploitation Detection} \label{subsec: att-det}
To demonstrate the capability of detecting malicious attacks, we simulated several malicious device and driver attacks which are illustrated in the section~\ref{subsec: threat}. We deploy our framework to the simulated exploitation environment to see whether it can detect the exploits or not. We reported our experiments in detecting software attacks and hardware attacks respectively.
\medskip

\noindent {\bf Detecting Software Attacks.} We instrumented the drivers of the five listed devices in Table~\ref{Tab: Dev} to simulate driver attacks targeting on the devices. We modified these drivers to issue malicious driver events which violate the system properties, causing device hangs or crashes. The voilated system properties and the consequence of corresponded violations are summarized in Table~\ref{Tab: Property}. In the experiment, our framework could successfully catch the invalid inputs from the instrumented drivers.

%\begin{table*}[t]
%\centering
% \scriptsize
%\caption{Summary of Software Attack Injection}\label{Tab: Property}
%\begin{tabular}{c|l|c} \hline
%\tabincell{c}{\bf Driver} & \tabincell{c}{\bf Property}& \tabincell{c}{\bf Consequence } \\\hline
%rtl8139 & \tabincell{l}{ The driver should not start new transaction  \\ when another transaction is in progress} & Device hangs \\\hline
%eepro100 & \tabincell{l}{The driver should not issue START \\ while  the device is working already} & \tabincell{c}{Device or \\ even system hangs} \\\hline
%e1000 & \tabincell{l}{The driver should not issue command \\ when MDIC is not clear}& Device hangs\\\hline
%bcm5751 &\tabincell{l}{The driver should not start new  EEPROM  \\ transaction when previous update is not finished}  & Device hangs \\\hline
%x520 & \tabincell{l}{The driver should not issue command \\ when MDIC is not clear}& Device hangs\\\hline
%\end{tabular}
%\end{table*}

\medskip
\noindent {\bf Detecting Hardware Attacks.} In this experiment, we mainly simulated two types of hardware attacks from malicious devices which are  illustrated in section~\ref{subsec: threat}: (1) the exploitation of buffer overflow vulnerability in the driver; (2) the code injection attack through DMA. In order to simulate malicious devices, we employ an open-source widely-used virtual machine, named QEMU~\cite{Bellard05}.  QEMU provides a rich set of virtual devices simulating hardware device behaviors. We modified the corresponded virtual devices of the five ones in Table~\ref{Tab: Dev} by introducing malicious behaviors as needed. We carried out our experiments on the QEMU virtual machine where the drivers and virtual devices are running under the monitor of our framework. The experiment details of the attacks are as follows.
\medskip
\begin{itemize}
\item {\bf Buffer Overflow Exploitation.} Since this attack requires buffer overflow vulnerability existing in device drivers, for each driver of the five devices, we injected a simple {\tt memcpy} based operation where the size and original buffer are controlled by the device. We instrumented the virtual devices to provide  the malformed inputs through device interface registers, causing buffer overflow in the modified drivers. The result show that our framework can successfully detect the malformed values of the interface registers through device checking.
\medskip
\item {\bf Code Injection Attack via DMA.} We instrumented the virtual devices to implement the example attack described in Figure~\ref{Fig: Att}. The malicious virtual devices could successfully execute the injected code and complete privilege escalation, e.g., creating a ``sudo'' user account. Our framework also detected this attack in a timely manner, as long as the injected code was placed in the device interface registers, the device checking reported the inconsistencies between the virtual device and FDM, discovering the attack.
\end{itemize}
\medskip

\subsection{Performance} \label{subsec: performance}
We evaluated the overhead introduced by HW/SW co-monitirng by measuring CPU and memory usages under four test scenarios. These test scenarios are common use cases of an Ethernet adapter, including ``{\em load driver}'', ``{\em scp files}'', ``{\em reset device}'', and a test suite which is a set of small test cases specifically designed for testing Ethernet adapters, including ``{\em sending a UDP/TCP packet}", ``{\em configuring IP addresses}", ``{\em ping for 30 seconds}", and so on.

We compared the CPU and memory usages under two system configurations: the native system without co-monitoring, denoted ``NAT.'', and the monitored system with co-monitoring, denoted as ``MON.".  Figure~\ref{Fig:PER} shows the results, where the usages in NAT. are normalized to 1. The results show that under most of the test cases, our co-monitoring approach introduces reasonable overhead, CPU overhead is 1.2x - 1.6x while memory overhead is 1.1x - 1.4x.

\medskip
\noindent{\bf Discussion}. We noticed that under some heavey workloads, the overhead is relatively high comparing to the normal situations. The reason is that these workloads caused intensive device/driver requests. Since our
framework symbolically executes the FDM upon every driver request, intensive driver requests will lead to signicant overhead incurred by symbolic execution. A potential solution is ``caching'' the driver request and FDM state transitions. We observed that in the scenarios with heavy traffics such as network packet transfering, same sequences of driver requests and FDM state transitions are often generated. To this end, we can cache the driver request and FDM state transitions explored by symbolic execution of the FDM. If encountering the same driver request and the FDM state, SEE directly reuse the cached FDM state transitions without invoking symbolic execution. As a result, the runtime overhead could be further mitigated. We plan to implement this optimization in our future work.
%\subsection{Discussion}

Our evaluation demonstrates that (1) this framework is useful: it can discover runtime errors in the device and the driver and (2) the framework is efficient: its performance overhead is modest. Overall, these results indicate that our approach has major potential in helping validate and debug device/driver together.

\section{Related Work} \label{Sec: related}
%Lei, et al.~\cite{Lei13, Lei14} proposed post-silicon conformance checking, an approach to checking if a silicon device conforms to its virtual device. It captures silicon device traces at runtime and checks the conformance with the virtual device by processing the captured traces offline.  We apply this technique to check if the device conforms to the FDM at runtime.  Our approach has two major improvements: (1) HW/SW co-monitoring is an on-line approach, monitoring device/driver interactions while the system is running; (2) we utilize a FDM as a reference model abstracting unnecessary details, which makes runtime checking efficient.
%
%%(2) Our approach is designed for the product release stage, focusing more on detecting hardware transient errors and malicious attacks rather than discovering logic errors;
%Li, et al.~\cite{Li2010} presented an automata-theoretic approach to HW/SW co-verification. It models HW/SW compositions as BPDS and verifies system properties through reachability analysis. Our approach conducts runtime HW/SW co-verification over the BPDS and has two major advantages: (1) we leverage concolic execution to explore the BPDS state space, avoiding state space explosions; (2) The bugs discovered by co-monitoring is the real bugs which occur at runtime. In contrast, static HW/SW co-verification might produce false positives as it is not carried out in the real environment (e.g., real devices and operating systems).
%
%Besides the above two closely related work.
Our work is related to the work in three areas: (1) driver verification, testing, and monitoring; (2) device validation, testing, monitoring; (3) HW/SW co-validation and co-verification. We discuss the corresponded related work below. \medskip

\noindent {\bf Driver validation and monitoring.}  A notable driver verification framework is Static Driver Verifier (SDV)~\cite{Ball06}, which focuses on verifying Windows Kernel API usages of Windows drivers. It applies software model checking to verify the system properties to which Windows drivers should conform. There has been much research focusing on exploring possible driver execution paths and test case generations~\cite{DDT10,Chipounov2011, SymDrive}, by leveraging runtime symbolic execution. Monitoring a runtime system is an alternative approach to enhancing system reliabilities. Nooks \cite{Swift05} and  Carburizer~\cite{Kadav09} detect driver errors at runtime and carry out simple mechanisms to handle the errors and recover the system. \medskip

\noindent {\bf Device validation and monitoring.} Device testing and validation are usually carried out in the post-silicon stage. Post-silicon validation is performed on silicon prototypes and testing devices. A significant amount of research has been focused on detecting and localizing bugs in silicon chips. A major difficulty of post-silicon bug detection and localization is the limited observability of silicon device internals.  Previous work \cite{Abramovici06, Park08} has developed hardware on-chip monitors to collect hardware execution traces with internal signals. Assertion-based verification~\cite{Boule06, Hu03, Jos03} and formal methods~\cite{backspace} have been used to analyze and debug the execution traces from on-chip monitors. Our approach also works on detecting and troubleshooting post-silicon bugs. Instead of validating internal implementations of hardware devices, we focus on monitoring HW/SW interfaces.\medskip

\noindent {\bf HW/SW interface assurance.}
There has been several research on HW/SW interface assurance in the validation stage. HW/SW co-verification and co-simulation are two major techniques. HW/SW co-verification mainly uses model checking~\cite{fmsd02} as well as symbolic execution ~\cite{Horn2013} which verify HW/SW interface protocols against the device and driver models. Research on co-simulation~\cite{GB95, HKM01, Ro94} explores the design space of HW/SW boundaries with hardware models. Both HW/SW co-verification and co-simulation operate against hardware designs and models rather than the implementation of the HW/SW interfaces. And neither of them is conducted in the real runtime environment. Therefore, how to eliminate false positives and reproduce the detected bugs are often challenging. Our approach is conducted on the real devices and drivers, all the detected bugs are real bugs occurred at runtime. Moreover, the malicious attacks or hardware transient errors can only be detected by HW/SW co-monitoring, as they only occur at deployment stage.

\section{Conclusions and Future Work} \label{Sec: con}
We have presented HW/SW co-monitoring, a co-verification approach to discovering bugs, failures, and malicious exploits during device/driver interactions. We evaluated our approach and discoved 9 bugs in commercial devices and drivers which have been deployed into production for years. The performance evaluation indicated that the introduced overhead is reasonable. These results demonstrate that HW/SW co-monitoring is effective and efficient in detecting logic bugs as well as malicious exploits over HW/SW interfaces at runtime.

For future work, we will explore two directions. First, we will implement our ``caching'' strategy to further mitigate the runtime overhead. Second, HW/SW co-monitoring provides a foundation for protecting computer systems at runtime by effectively detecting the failures and exploits in HW/SW interfaces. In future, we will research how to efficiently prevent the failures from affecting the rest of the systems by either rectifying or isolating the discovered failures.

%Besides the future work discussed in Section~\ref{sec: dis}, our immediate next step work will focus on two directions: (1) evaluating more hardware devices, not only Ethernet adapters, but also other categories such as USB devices and wireless devices; (2) investigating more malicious attack scenarios which involve hardware and software interactions and evaluating whether our approach can detect them.

\bibliographystyle{IEEEtran}
\bibliography{ddcom}

\end{document}